\documentclass[preprint,pra,showpacs,showkeys]{revtex4}
\usepackage{graphicx,amsmath,amsfonts,amssymb}
\usepackage{times}
\begin{document}

\title{\Large A Question: Quantum Critical Phenomena in XY Spin-Chain Systems With Dzyaloshinskii-Moriya Interactions }

\author{Chuan-Jia Shan\footnote{E-mail: shanchuanjia1122@yahoo.com.cn}}
\author{Wei-Wen Cheng}
\author{Ji-Bing Liu}
\author{Tang-Kun Liu}
\author{Yan-Xia Huang}
\author{Hong Li}
\affiliation{College of Physics and Electronic Science, Hubei Normal
University, Huangshi 435002, China}

\date{\today}

\begin{abstract}
\indent Jordan-Wigner transformation and Bogolyubov transformation
are the main steps of the diagonalization of Hamiltonian and paly an
important role in the statistical mechanics calculations for
one-dimensional Heisenberg spin chain model. Many methods can be
exploited as a tool to detect quantum phase transition, regions of
criticality and scaling behavior in the vicinity of a quantum phase
transition, such as geometric phase, fidelity susceptibility, order
parameter, and entanglement entropy, which have direct relation with
Bogolyubov transformation. We diagonalized the Hamiltonian in XY
spin-chain systems with Dzyaloshinskii-Moriya interactions, the
results shows that only the energy spectrum but not the coefficients
of the Bogolyubov transformation depends on DM interaction.
Therefore, the DM interaction may not influence the critical
magnetic field of quantum phase transitions and not induce new
critical regions in the XY spin model. Moreover, we further prove
the ideas by the methods of geometric phases in this model.
\end{abstract}



\maketitle

\renewcommand{\baselinestretch}{1.5} \large\normalsize

\setlength{\parindent}{2em} Quantum phase transition(QPT) in a
many-body system, which occurs at absolute zero temperature and is
purely driven by quantum fluctuations, is the structural change in
the properties of the ground state.  The associated level crossings
lead to the presence of non-analyticities in the energy spectrum.
Indeed, it has been noted that quantum phase transitions present in
condensed matter systems can be described from the view of geometric
phase[1], fidelity susceptibility[2], and entanglement[3]. The
Dzyaloshinskii-Moriya interaction is present in many low-dimensional
materials. Such antisymmetric exchange interaction take places due
to spin-orbit coupling, this interaction has a number of important
consequences and may cause a number of unconventional phenomena. In
this paper we analyze the effect of the Dzyaloshinskii- Moriya
interaction on the Bogolyubov transformation and the energy spectrum
of quantum spin chains.

The model we study is a physical  XY spin chain consists of N spins
with nearest-neighbor interactions and an external magnetic field.
In the presence of Dzyaloshinskii-Moriya interactions, the
Hamiltonian of the system reads
\begin{eqnarray}
H=-\sum_{j=1}^{N}(J\frac{1+\gamma }{2}\sigma _{j}^{x}\sigma
_{j+1}^{x}+J\frac{1-\gamma }{2}\sigma _{j}^{y}\sigma
_{j+1}^{y}+\frac{D}{2}(\sigma _{j}^{x}\sigma _{j+1}^{y}-\sigma
_{j}^{y}\sigma _{j+1}^{x})+\lambda \sigma _{j}^{z}),
\end{eqnarray}
where $\lambda$ is the strength of the external magnetic field on
every site, D is the z component of the Dzyaloshinskii-Moriya
interaction, $\sigma^{x,y,z}$ are the Pauli matrices, and $N$ is the
number of sites. We assume periodic boundary conditions which
satisfy $\sigma _{N+1}^{x}=\sigma _{1}^{x}, \sigma _{N+1}^{y}=\sigma
_{1}^{y}, \sigma _{N+1}^{z}=\sigma _{1}^{z}$.

 A major advantage of
the above Hamiltonian is that it can be diagonalized exactly. Let us
briefly review the main steps of the diagonalization of H, the set
of n qubits in Eq. (1) can be mapped to a system of n spinless
fermions via the Jordan-Wigner transformation. Let us define the
raising and lowing operators $a_{i}^{+}$,$a_{i}^{-}$ and introduce
Fermi operators
 $c_{j}^{+}$,$c_{j}$,\\
 \begin{eqnarray}a_{i}^{+}=\frac{1}{2}(\sigma _{i}^{x}+i\sigma
_{i}^{y})=\exp (i\pi \sum_{j=1}^{i-1}c_{j}^{+}c_{j})c_{i}, \\
a_{i}^{-}=\frac{1}{2}(\sigma _{i}^{x}-i\sigma _{i}^{y})=\exp (-i\pi
\sum_{j=1}^{i-1}c_{j}^{+}c_{j})c_{i}^{+}.
\end{eqnarray}
So that, the Hamiltonian (1) can be transformed into
\begin{eqnarray}
H=-\sum_{i=1}^{N}[(J+iD)c_{i}^{+}c_{i+1}+(J-iD)c_{i+1}^{+}c_{i}+J\gamma(c_{i}^{+}c_{i+1}^{+}+c_{i+1}c_{i})+
\lambda(1-2c_{i}^{+}c_{i})]
\end{eqnarray}
After the Fourier transformation

\begin{eqnarray}
d_{k}=\frac{1}{\sqrt{N}}\sum_{j}c_{j}\exp (\frac{-i2\pi jk}{N}),
   k=-\frac{N-1}{2},\cdots , \frac{N-1}{2}
\end{eqnarray}
Introducing Bogoliubov transformation
\begin{eqnarray}
d_{k}=c_{k}\cos
\frac{\theta _{k}}{2}-ic_{-k}^{\dagger}\sin \frac{\theta _{k}}{2}
\end{eqnarray}
with
\begin{eqnarray}
\cos \theta _{k}=\frac{\lambda-J\cos(\frac{2\pi
k}{N})}{\sqrt{(\lambda-J\cos(\frac{2\pi k}{N}))^{2}+ J^{2}\gamma
^{2}\sin ^{2}(\frac{2\pi k}{N})}}\\
\sin \theta _{k}=\frac{J\gamma\sin(\frac{2\pi
k}{N})}{\sqrt{(\lambda-J\cos(\frac{2\pi k}{N}))^{2}+ J^{2}\gamma
^{2}\sin ^{2}(\frac{2\pi k}{N})}}
\end{eqnarray}

 the Hamiltonian [1] becomes
\begin{eqnarray}
H=\sum_{j}\Lambda _{k}(d_{k}^{\dagger }d_{k}-\frac{1}{2})
\end{eqnarray}
where $\Lambda _{k}=2\sqrt{(\lambda-J\cos(\frac{2\pi k}{N}))^{2}+
J^{2}\gamma ^{2}\sin ^{2}(\frac{2\pi k}{N})}+2D\sin (\frac{2\pi
k}{N})$

It should be noted here that only the energy spectrum $\Lambda _{k}$
but not the coefficients of the Bogolyubov transformation $\sin
\frac{\theta _{k}}{2}, \cos \frac{\theta _{k}}{2}$ depends on D,
which is consistent with the results in Ref.[4], but has some
differences with the results in Ref.[5]. In Ref.[1], the author
shows that the geometric phase of the ground state in the XY model
obeys scaling behavior in the vicinity of a quantum phase
transition. The established relation between the geometric phase and
quantum phase transitions is a very general result of many-body
systems. We can introduce the DM interaction and work out the
behavior of this model, the DM interaction can not influence the
critical magnetic field of quantum phase transitions and not induce
new critical regions in the XY spin model.


\begin{thebibliography}{1}
\bibitem{1}S. L. Zhu, Phys. Rev. Lett. 96, 077206 (2006).
\bibitem{2}P. Zanardi and N. Paunkovic, Phys. Rev. E 74, 031123
(2006). P. Zanardi, M. Cozzini, P. Giorda, J. Stat. Mech. 2, L02002
(2007). P. Zanardi, H. T. Quan, X. G. Wang, and C. P. Sun, Phys.
Rev. A 75, 032109 (2007).
\bibitem{3}A.Osterloh, Luigi Amico, G. Falci, and Rosario Fazio,
Nature 416, 608 (2002)
\bibitem{4}Oleg Derzhko, Taras Verkholyak, Taras Krokhmalskii, and Helmut B¨¹ttner Phys. Rev. B 73, 214407
(2006)\\
W.W Cheng and J.M Liu Phys. Rev. A (in press)
\bibitem{5}Y.C Li and S.S Li
Phys. Rev. A 79, 032338 (2009)

\end{thebibliography}
\end{document}